\documentstyle[11pt]{article}
\begin{document}
\title{{\normalsize{{\hskip 9cm} BIHEP-TH-97-12  , 1997}}\\
 Futher Study of CP Violation and Branching Ratios 
 for $\bar{B^0}\to J/\psi K_s$ and
 $\bar{B^0}\to \phi K_s$ in the Standard Model and Beyond 
 \thanks{Supported in part by National Natural Science Foundation
          of China}}

\author{Dong-Sheng Du and Mao-Zhi Yang\\
    CCAST(WORLD LABORATORY)P.O.BOX 8730,BEIJING,100080\\     
                            and \\
    Institute of High Energy Physics, Chinese Academy of Sciences,\\
  P.O.Box 918(4), Beijing, 100039,People's Republic of China
         \thanks{Mailing address}}
\date{}
\maketitle

\vskip 0.3in
\begin{abstract}
In this work we study the CP violation for $\bar{B^0}\to J/\psi K_s$ and
$\bar{B^0}\to \phi K_s$ up to
leading and next-to-leading order QCD corrections in the standard model,
two-Higgs-doublet model and the minimal supersymmetric extension of the
standard model. We also study the effect of new physics on the
branching ratios of these two decay modes. We find that within the
parameter space constrained by the observation of the decay $b\to s\gamma$,
 new physics does not affect the CP
asymmetries greatly, and the prediction of new physics to the branching
ratios of $\bar{B^0}\to J/\psi K_s$ and $\bar{B^0}\to \phi K_s$ is the same
as that of the standard model up to a minor discrepancy as far as the Yukama
couplying constants are perturbative.
 
\end{abstract}
\newpage

\begin{center}
\section*{I.INTRODUCTION}
\end{center}

  The decay of $\bar{B^0}\to J/\psi K_s$ is expected to be one of the most
promising channels to study the CP violation in the B decays at the B
factory. In the standard model CP violation is produced via the 
complex phase of the Cabibbo-Kobayashi-Maskawa (CKM) matrix. Because the 
CKM matrix is unitary there is the unitarity equation
\begin{equation}
 \sum_i V_{ij} V_{ik}^*=0~~(j\not= k). 
 \end{equation}
This equation can be represented as a closed triangle in the complex plane,
which is called the unitarity triangle. Since the most poorly known entries
in the CKM matrix are contained in the triangle equation
\begin{equation}
V_{ud} V_{ub}^*+V_{cd} V_{cb}^*+V_{td} V_{tb}^*=0, 
\end{equation}
so this equation is most useful from  the phenomenological point of view
\cite{1}.
This equation can be visualized as the unitarity triangle in Fig.1. The angles
of $\alpha$, $\beta$, and $\gamma$ in Fig.1 can be determined through the
detection of the CP violation of the decay modes $\bar{B}\to \pi\pi$, 
$\bar{B}\to \psi K_s$, $\bar{B}\to\phi K_s$, and $B_s\to \rho^0 K_s$ 
respectively\cite{1,2}. In the standard model CP violation is only produced 
through the CKM mechanism. It is believed that if the sum of the three angles
$\alpha$, $\beta$, and $\gamma$ is measured to be $\pi$, the standard model 
gives the correct CP violating mechanism. If not, then there must be new 
physics beyond the standard model.

 The decay $\bar{B^0}\to J/\psi K_s$ and $\bar{B^0}\to \phi K_s$ are
based on the transition $b\to q\bar{q}s$ in the quack level, where $q$ can
be $c$ and $s$. Their CP conjugate processes are
$B^0\to J/\psi K_s$ and $B^0\to \phi K_s$ respectively. Thus both the state
and CP conjugate state of the
initial B meson can decay into the same final state. It's CP violating
parameter, CP asymmetry, can be described by\cite{3}
  \begin{eqnarray}
    {\cal A}_{cp}&=&\displaystyle\frac{\int_0^{\infty}
           [\Gamma(B_{phys}^0 (t)\rightarrow f)-
              \Gamma(\bar B_{phys}^0 (t)\rightarrow f)]dt
             }{\int_0^{\infty}
           [\Gamma(B_{phys}^0 (t)\rightarrow f)+
              \Gamma(\bar B_{phys}^0 (t)\rightarrow f)]dt} \nonumber\\[4mm]
      &=&\displaystyle\frac{1-|\xi|^2-2Im\xi(\Delta m_B/\Gamma_B)}
      {(1+|\xi|^2)[1+(\Delta m_B/\Gamma_B)^2]},
\end{eqnarray}
where $\xi=\pm\frac{q}{p}\frac{\bar{A}}{A}$, here `+' for CP even $|f\rangle$,
$`-'$ for CP odd $|f\rangle$ \cite{du}.
For the present two decay modes that we study, 
$\xi_f=-\left(\frac{q}{p}\right)_B
                           \left(\frac{q}{p}\right)_K^* \frac{\bar{A}}{A}$.
The parameters $p$, $q$ are defined in the physical state of $B$ and $K$ 
mesons. The $K_s$ meson is defined as 
$|K_s\rangle=p_K|K^0\rangle+q_K|\bar{K}^0\rangle$, and the physical state 
of the neutral B meson is defined as
 \begin{eqnarray}
 |B_L\rangle &=&p_B|B^0\rangle +q_B|\bar{B}^0\rangle,\nonumber\\[4mm]
 |B_H\rangle &=&p_B|B^0\rangle -q_B|\bar{B}^0\rangle,
 \end{eqnarray} 
 \noindent where L, H denote light and heavy respectively.
In the standard model the decay $\bar{B^0}\to J/\psi K_s$ is dominated by the 
tree level contribution, and $\xi_{\psi K_s} \simeq -e^{-2i\beta}$ under 
the tree level approximation. The $\bar{B^0}\to \phi K_s$ is forbidden
at the tree level, its relevant value of $\xi$ is also equal to $-e^{-2i\beta}$
approximately. However, in some non-standard models such as
the two-Higgs-doublet model (THDM) and the minimal supersymmetric 
extension of the standard model, CP violation can either be produced through the CKM mechanism, or through
the Higgs sector, or from both. So in order to investigate the very nature
of CP violation, the B decays should be studied
carefully in both standard and non-standard models.

  In this work we study what impact the new mechanisms of
non-standard model can give to the measurement of the CP violation of
$\bar{B^0}\to J/\psi K_s$ and $\bar{B^0}\to \phi K_s$. To take into account
the nonfactorization corrections, the inverse of the color number 
$\frac{1}{N_c}$ should be replaced with $\frac{1}{N_c}+\chi$, where $\chi$
stands for the nonfactorization corrections, and this replacement amounts
to changing the color number from $N_c=3$ \cite{cheng}.  Here we consider three conditions
with $N_c=2$, 3 and $\infty$. We also investigate
the new physics effects on the branching ratios of $\bar{B^0}\to J/\psi K_s$
and $\bar{B^0}\to \phi K_s$. The paper is organized as the following. In
section II we study the CP violation and the branching ratio of 
$\bar{B^0}\to J/\psi K_s$ in the SM,
THDM amd the minimal supersymmetric extension of SM (MSSM). Section III
is devoted to the study of $\bar{B^0}\to \phi K_s$ in those three models. 
Section IV is for the discussion and conclusion.

\begin{center}
\section*{II.The study of $\bar{B^0}\to J/\psi K_s$}
\end{center}

\begin{center}
1) {\bf CP violation of  
     $\bar{B^0}\to J/\psi K_s$ in the standard model}
\end{center}

The low energy effective Hamiltonian relevant to our study is\cite{4,5}

\begin{eqnarray}
{\cal H}_{eff}&=&\frac{G_F}{\sqrt{2}}\left[\displaystyle\sum_{q=u,c}v_q
\left\{ \frac{}{}
 Q_1^qC_1(\mu)+Q_2^q C_2(\mu)\right.\right.\nonumber\\[4mm]
&&+\left.\left.\displaystyle\sum_{k=3}^{10} Q_k C_k(\mu)\right\}\right]+
 {\cal H.C.} ,
\end{eqnarray}
 where $C_k(\mu)$ (k=1, $\cdots$, 10) are Wilson Coefficients (WC) which are
 calculated in the renormalization group improved perturbation
 theory and include
 leading and next-to-leading order QCD corrections. $v_q$ is the product of
 Cabibbo-Kobayashi-Maskawa (CKM) matrix elements and defined as
 $$
 v_q=\left\{
           \begin{array}{rl}
            v_{qd}^*v_{qb}  & b\to d~~transitions,\\
            v_{qs}^*v_{qb}  & b\to s~~transitions.
           \end{array}\right.
 $$   
 The ten operators are taken in the following form\cite{4}:
\begin{eqnarray}
   Q_1^u&=&(\bar{u}_{\alpha}b_{\beta})_{V-A}(\bar{q}_{\beta}u_{\alpha})_{V-A},
    ~~~~~~~~~
   Q_2^u=(\bar{u}b)_{V-A}(\bar{q}u)_{V-A},\nonumber\\
   Q_3&=&(\bar{q}b)_{V-A}\displaystyle\sum_{q'}(\bar{q'}q')_{V-A},~~~~~~~~~~~~
   Q_4=(\bar{q}_{\beta}b_{\alpha})_{V-A}             
        \displaystyle\sum_{q'}(\bar{q'}_{\alpha}q'_{\beta})_{V-A}, \nonumber \\
   Q_5&=&(\bar{q}b)_{V-A}\displaystyle\sum_{q'}(\bar{q'}q')_{V+A},~~~~~~~~~~ 
   Q_6=(\bar{q}_{\beta}b_{\alpha})_{V-A}             
        \displaystyle\sum_{q'}(\bar{q'}_{\alpha}q'_{\beta})_{V+A},  \\
   Q_7&=&\frac{3}{2}(\bar{q}b)_{V-A}\displaystyle\sum_{q'}e_{q'}
    (\bar{q'}q')_{V+A},~~~~~~~~~
   Q_8=\frac{3}{2}(\bar{q}_{\beta}b_{\alpha})_{V-A}                
   \displaystyle\sum_{q'}e_{q'}(\bar{q'}_{\alpha}q'_{\beta})_{V+A}, \nonumber\\
 Q_9&=&\frac{3}{2}(\bar{q}b)_{V-A}\displaystyle\sum_{q'}e_{q'}(\bar{q'}q')_{V-A},
   ~~~~~~~~~~
   Q_{10}=\frac{3}{2}(\bar{q}_{\beta}b_{\alpha})_{V-A}            
\displaystyle\sum_{q'}e_{q'}(\bar{q'}_{\alpha}q'_{\beta})_{V-A},\nonumber
 \end{eqnarray}
\noindent where  $Q_1^q$ and $Q_2^q$ are current-current operators, q=u, c.
For $q=c$ case, $Q_1^c$ and $Q_2^c$ are obtained through making substitution
$u\to c$ in  $Q_1^u$ and $Q_2^u$.
$Q_3\sim Q_6$ are QCD penguin operators, the sum $\displaystyle\sum_{q'}$ is
runing over all flavors being active at $\mu=m_b$ scale,
$q'=\{u, d, s, c, b\}$. 
$Q_7\sim Q_{10}$ are electroweak penguin operators, $e_{q'}$ are the electric
charges of the relevant quarks in unit e which is the charge of the proton. 
The subscripts $\alpha,~
\beta$ are $SU(3)_c$ color indices. $(V\pm A)$ referes to $\gamma_{\mu}(1\pm
\gamma_5)$.                                             
  
  For our calculation  we use the renormalization scheme independent form of 
the Wilson Coefficients $C_i'(\mu)$,
   \begin{eqnarray}
  C'_1&=&\bar{C}_1,~~~~~~~~~~~~~C'_2~=~\bar{C}_2,\nonumber\\
  C'_3&=&\bar{C}_3-P_s/N_c,~~~~~~C'_4~=~\bar{C}_4+P_s,\nonumber\\
  C'_5&=&\bar{C}_5-P_s/N_c,~~~~~~C'_6~=~\bar{C}_6+P_s,\\
  C'_7&=&\bar{C}_7+P_e,~~~~~~~~~C'_8~=~\bar{C}_8,      \nonumber\\
  C'_9&=&\bar{C}_9+P_e,~~~~~~~~~C'_{10}~=~\bar{C}_{10},\nonumber
  \end{eqnarray} 
where $\bar{C}_i(\mu)$ is\cite{5,6,7}
\begin{equation}
{\bf \bar{C}}(\mu)=(\hat{1}+\hat{r}_s^T\alpha_s(\mu)/4\pi+
                        \hat{r}_e^T\alpha_{em}(\mu)/4\pi)\cdot 
                        {\bf C}(\mu),
\end{equation}
and 
 \begin{eqnarray*}  
  P_s&=&\displaystyle\frac{\alpha_s}{8\pi}\bar{C_2}(\mu)
       \left[\frac{10}{9}-G(m_q,q,\mu)\right] , \\[4mm]
  P_e&=&\displaystyle\frac{\alpha_{em}}{3\pi}\left(\bar{C_1}(\mu)+
  \displaystyle\frac{\bar{C_2}(\mu)}{Nc}\right)\left[\frac{10}{9}-
 G(m_q,q,\mu)\right],\\[4mm]
 G(m_q,q,\mu)&=&-4\int_0^1 x(1-x)dxln\displaystyle\frac{m_q^2-x(1-x)q^2}
{\mu^2},
\end{eqnarray*} 
 \noindent here q=u,c. The renormalization scheme independent Wilson 
 coefficients $\bar{C}_i(\mu)$ has been given in Ref. \cite{8},
  \begin{eqnarray}
  \bar{C_1}&=&-0.313,~~~~\bar{C_2}~=~1.150,~~~~\bar{C_3}~=~0.017,\nonumber\\
 \bar{C_4}&=&-0.037,~~~~\bar{C_5}~=~0.010,~~~~\bar{C_6}~=~-0.046, \nonumber\\
 \bar{C_7}&=&-0.001\cdot\alpha,~~~
 \bar{C_8}~=~0.049\cdot\alpha, ~~~
 \bar{C_9}~=~-1.321\cdot\alpha,   \\
 \bar{C_{10}}&=&0.267\cdot\alpha.\nonumber
 \end{eqnarray}
 
 With the effective Hamiltonian given in eq.(5) we obtain
 \begin{eqnarray}
 \bar{A}&\equiv &\langle J/\psi K_s|H_{eff}|\bar{B}^0\rangle=
        q_K^*\langle J/\psi \bar{K}^0|H_{eff}|\bar{B}^0\rangle\nonumber
	\nonumber\\[4mm]
        &=&q_K^*\displaystyle\frac{G_f}{\sqrt{2}}
          \left[(C_1'+\frac{C_2'}{N_c})v_c+\displaystyle\sum_{q=u,c}
          v_q\left[C_3'+\frac{C'_4}{N_c}+C_5'+\frac{C'_6}{N_c}+ 
          \frac{3}{2}e_c(C_7'+\frac{C'_8}{N_c}+C_9'+\frac{C'_{10}}{N_c})
          \right]\right]\nonumber\\[4mm]
         &&~ \cdot\langle J/\psi|(\bar{c}c)_{V-A}|0\rangle\langle
          \bar{K}^0|(\bar{s}b)_{V-A}|\bar{B}^0\rangle,
	  \end{eqnarray} 
 After we get eq.(10), we are now in the position to calculate 
 $\xi_{\psi K_s}$,
 \begin{equation}
 \xi_{\psi K_s}=-\left(\frac{q}{p}\right)_B\left(\frac{q}{p}\right)_K^* 
                          \frac{\bar{A}}{A}, \end{equation}
\noindent and
\begin{eqnarray*}
 \left(\frac{q}{p}\right)_B\left(\frac{q}{p}\right)_K^*&\simeq&
       \displaystyle\frac{v_{tb}^*v_{td}}{v_{tb}v_{td}^*}\cdot
        \displaystyle\frac{v_{cs}^*v_{cd}}{v_{cs}v_{cd}^*}\\[4mm]
        &\simeq&\displaystyle\frac{1-\rho-i\eta}{1-\rho+i\eta}\\[4mm]
        &\simeq&e^{-2i\beta},\end{eqnarray*} 
\noindent where $\rho$, $\eta$ are the parameters 
in the Wolfenstein parameterization
for the CKM matrix. Substitute the numerical values of $C'_i$ into eq.(10) and
(11), we can get the numerical result of $\xi_{\psi K_s}$. For the case
without loop correction, only $C_1'$ and $C_2'$ contribute,
\begin{equation}
\xi_{\psi K_s}\approx -e^{-2i\beta}\approx -\frac{1-\rho-i\eta}{1-\rho+i\eta}.
 \end{equation}
\noindent For the case with loop corrections 
all $C_1'\cdots C_{10}'$ contribute,  
 \begin{eqnarray}
\xi_{\psi K_s}&=&
 -e^{-2i\beta}\displaystyle\frac{118.73-\rho+i\eta}{118.73-\rho-i\eta}
    \nonumber\\[4mm]
 &\stackrel{or}{=}&-\displaystyle\frac{1-\rho-i\eta}{1-\rho+i\eta}
             \displaystyle\frac{118.73-\rho+i\eta}{118.73-\rho-i\eta}.
  \end{eqnarray} 
\noindent Experimental results have
constrained $\rho$ and $\eta$ into a small area in
the $\rho$-$\eta$ plane\cite{1}, approximately, $-0.2<\rho<0.3$, and $0.2<\eta<0.4$.
From eq.(12) and (13) we can see that loop diagram correction to
$\xi_{\psi K_s}$ is only in the order of ${\cal O}(10^{-3})$.
So the unitarity triangle angle $\beta$
can be determined through the measurement of the CP asymmetry in
$\bar{B^0}\to J/\psi K_s$ up to an approximation of the order
${\cal O}(10^{-3})$. The eq.(13) is obtained with $N_c=3$, when $N_c=2$, 
$\infty$
the loop corrections are also in the same order of ${\cal O}(10^{-3})$.
This conclusion has been obtained through an estimate that 
the penguin contributions are propotional to
$V_{tb}V_{ts}^*\simeq \lambda^2$, $V_{cb}V_{cs}^*\simeq \lambda^2$
and $V_{ub}V_{us}^*\simeq \lambda^4$ ($\lambda\approx 0.22 $), so
up to very small corretions the penguin contributions
have the same weak phase as the tree diagram contribution, then the penguin
contribution affects the CP violation extremely small \cite{1,Bigi}. Our 
calculation using the QCD corrected Hamiltanian confirms this conclusion.

\begin{center}
{\bf 2) CP violation of $\bar{B^0}\to J/\psi K_s$ in the 
        two-Higgs-doublet model}
\end{center}
        
  In the two-Higgs-doublet model\cite{9} the diagrams contributing to
 $\bar{B^0}\to J/\psi K_s$ are shown in Fig.2.
 The relevant low energy effective Hamiltonian has the same form as eq.(5),
 but the Wilson coefficients at the scale $\mu=M_W$ contain an extra
 contribution from the charged Higgs loop diagrams. We calculated the new
 contributions at the scale $\mu=M_W$ at first, and then using the 
 two-loop-renormalization group equation to evolve them down to the scale
 $\mu=m_b$. The initial conditions in the scale $\mu=M_W$ are
 \begin{equation}
 C_i(M_W)=C_i^{SM}(M_W)+C_i^{H^\pm}(M_W), 
 \end{equation}
\noindent where the SM contribution $C_i^{SM}(M_W)$'s have already been 
 given in Ref. \cite{5}. In this paper, we 
 calculate the contributions of charged 
Higgs loop diagrams, the results are
\begin{eqnarray}
C_1^{H^\pm}(M_W)&=&0,\nonumber\\[4mm]
C_2^{H^\pm}(M_W)&=&0,\nonumber\\[4mm]
C_3^{H^\pm}(M_W)&=&-\displaystyle\frac{\alpha_s(M_W)}{24\pi}F_{H^\pm}(x'_t),
   \nonumber\\[4mm]
C_4^{H^\pm}(M_W)&=&\displaystyle\frac{\alpha_s(M_W)}{8\pi}F_{H^\pm}(x'_t),
   \nonumber\\[4mm]
C_5^{H^\pm}(M_W)&=&-\displaystyle\frac{\alpha_s(M_W)}{24\pi}F_{H^\pm}(x'_t),
           \\[4mm]
C_6^{H^\pm}(M_W)&=&\displaystyle\frac{\alpha_s(M_W)}{8\pi}F_{H^\pm}(x'_t),
  \nonumber\\[4mm]
C_7^{H^\pm}(M_W)&=&\displaystyle\frac{\alpha}{6\pi}\left[E_{H^\pm}(x'_t)-D_{H^\pm}(x'_t)
                                     \right],\nonumber\\[4mm]
C_8^{H^\pm}(M_W)&=&0,\nonumber\\[4mm]
C_9^{H^\pm}(M_W)&=&
   \displaystyle\frac{\alpha}{6\pi}\left[E_{H^\pm}(x'_t)-D_{H^\pm}(x'_t)
   +\displaystyle\frac{1}{2sin^2\theta_w}D_{H^\pm}(x'_t)\right],
   \nonumber\\[4mm]
C_{10}^{H^\pm}(M_W)&=&0,\nonumber
\end{eqnarray} 
\noindent where
   \begin{eqnarray}
   x'_t&=&\displaystyle\frac{m_t^2}{M^2_{H^\pm}},\nonumber\\[4mm]
   E_{H^\pm}(x)&=&\displaystyle\frac{F_u^2}{18}\left[\displaystyle\frac{38x-79x^2+47x^3}{6(1-x)^3}+
                  \displaystyle\frac{4x-6x^2+3x^4}{(1-x)^4}lnx\right],
		  \nonumber\\[4mm]
   F_{H^\pm}(x)&=&\displaystyle\frac{F_u^2}{6}\left[\displaystyle\frac{16x-29x^2+7x^3}{6(1-x)^3}+
                  \displaystyle\frac{2x-3x^2}{(1-x)^4}lnx\right],
		  \\[4mm]
   D_{H^\pm}(x)&=&-\displaystyle\frac{F_u^2}{2}x_t\left[\displaystyle\frac{x}{1-x}+
                 \displaystyle\frac{x}{(1-x)^2}lnx\right],\nonumber
 \end{eqnarray} 
 where $F_u$ is the coupling constant between the charged Higgs and the 
 up- and down-type quarks. We consider two distinct two-Higgs-doublet modes,
 model I and model II, which naturally avoid tree-level flavor changing 
 neutral current (FCNC). In model I, one doublet $(\phi_2)$ gives masses to 
 all fermions and the other doublet $(\phi_1)$ essentially decouples from
 the fermions, and here $F_u=cot\beta$. In model II, $\phi_2$ gives mass to
 the up-type quarks, while the other $(\phi_1)$ gives the down-type quarks 
 masses, and in this model $F_u=cot\beta$, where $tan\beta\equiv v_2/v_1$,
 which is the ratio of the vacuum expectation values (VEV) of the two 
 Higgs doublet.
 
 After presenting the initial values of the Wilson coefficients $C_i(M_W)$
 in the two-Higgs-doublet model, now we shall evolve them down to the scale
 $\mu=m_b$, so that to take into account the QCD corrections. The 
 renormalization-group equation for $C(\mu)$ is given by\cite{5},
\begin{equation}
\left[\mu\frac{\partial}{\partial \mu}+\beta(g)\frac{\partial}{\partial g}
   \right]C(\frac{M_W^2}{\mu^2},g^2,\alpha)=\hat{\gamma}^T(g^2,\alpha)
                                   C(\frac{M_W^2}{\mu^2},g^2,\alpha),
\end{equation}
\noindent where $\hat{\gamma} $ is the anomalous 
dimensions. A. Buras et al. have 
calculated the $10\times 10 $ anomalous dimension matrix involving 
current-current operators, QCD penguin operators, and electroweak penguin
operators up to next-to-leading order QCD corrections in the standard 
model\cite{4}. Now in the two-Higgs-doublet model and the supersymmetric extension
of the standard model, the anomalous dimension matrix $\hat{\gamma}$ is just 
the same as that in the standard model. In this section and the next we will
use the anomalous dimension matrix $\hat{\gamma} $ calculated by Buras et al. 
to evolve the Wilson coefficients from scale $\mu=M_W$ down to the $\mu=m_b$ 
scale.

 The solution of the renormalization-group equation (19) is given\cite{5} by
 \begin{equation}
 C(\mu)=\hat{U}(\mu,M_W,\alpha)C(M_W), 
 \end{equation}
 \noindent where $\hat{U}(\mu,M_W,\alpha)$ is 
 the evolution matrix from $M_W$ down to
 $\mu<M_W$. By resolving the eq.(17), we can get the numerical result for
 $\hat{U}(m_b,M_W,\alpha)$, which is presented in Appendix A.
  
 The difference between the phases of $B^0-\bar{B}^0$ mixing in the
 standard model and in the two-Higgs-doublet model is trivial\cite{10}. There
 are two types of box diagrams which contribute to the $B^0-\bar{B}^0$  
 mixing: the diagrams with one W and one charged Higgs propagating
 in it and the diagrams with two charged Higgses. The charged Higgs 
 contributions
 to the mixing involve the same CKM factors and almost have the same
 phase as the W-box contributions\cite{11,12}. So we will take 
 $(q/p)_B=\frac{v_{tb}^*v_{td}}{v_{tb}v_{td}^*}$, which is the same as in the
 standard model. The case of the $K^0-\bar{K}^0$ mixing is similar to the 
 $B^0-\bar{B}^0$ mixing. The contributions of charged Higgs bosons to the 
 $K^0-\bar{K}^0$ mixing present a negligible phase difference from the 
 standard model.
 
 With the Wilson coefficients calculated in the two-Higgs-doublet model, and
 using eq.(3), (10) and (11), we can get the results of the CP asymmetry of
$\bar{B^0}\to J/\psi K_s$. We now discuss our numerical results.

In the parameter space constrained by the observed rate of $b\to s\gamma$, the
possibility that charged-Higgs exchange induces large CP-violating effects
through the complex phase of $v_2/v_1$ has 
been ruled out\cite{10,13}. 
In the continuing work we will drop
the complex phase of the VEV, i.e., only take into account the cases that
$v_2/v_1$ is a real number.

  The observation of the decay $b\to s\gamma$ gives a correlated bound 
to $cot\beta$ and the charged Higgs mass $M_{H^\pm}$\cite{12}
\begin{equation}
|cot\beta|(100GeV/M_{H^\pm})<0.8. 
\end{equation}
We make our calculation within this constraint of $cot\beta$ and $M_{H^\pm}$,
and also keep $cot\beta <3$, which is the region of $cot\beta$ that keeps the
Yukawa couplying perturbative\cite{BHP}. We calculate three cases
with the number of color
$N_c=2,~3$ and $\infty$. The result is that within the 
parameter space constrained by the experiment, charged Higgs effect reduces
the CP asymmetry. But the effect is too small to distinguish the
two-Higgs-doublet model from the SM.

\begin{center}
{\bf 3) CP violation of $\bar{B}^0\to J/\psi K_s$ in the MSSM}
\end{center}

In the minimal supersymmetric extension of the standard model (MSSM), except 
for the usual particle fields in the standard model, 
there are the relevant supersymmetric partners, and
an extra Higgs doublet. They add the new contributions to the process of
$b\to c\bar{c}s$. They are the contributions from: i)the charged Higgs loop
diagrams, ii)the up-type squarks and chargino loop diagrams, iii) the 
down-type squarks, gluino and neutralino loop diagrams, which are shown in
Fig.3. For the basic structure of the minimal supersymmetric extension of 
the standard model, we refer the readers to the Refs. \cite{16,17}.

Because the assumption is quite strong that all the parameters in the soft
supersymmetric breaking terms, such as the masses and couplings of the MSSM
scalars and fermions, are unified at the grand unification (GUT) scale
$\mu=M_{GUT}$, we relax all the constraints to the parameters of MSSM from
this assumption and investigate the effects of the more general structure
of MSSM. So the free parameters are the following: the masses of the left and
right-handed up and down-type squarks, $m_{\tilde{fL}}$, $m_{\tilde{fR}}$;
the mixing angles of the right and left-handed 
squarks, $\alpha_{\tilde{f}}$; the wino and charged Higgs masses
$M_{\tilde{W}}$ and $M_{H^\pm}$; the superpotential $\mu$ and $cot\beta$.
For simplicity the following features should be assumed:\\
(i) Supersymmetric loop diagram contributions to $B\to X_s\gamma$ 
 decay are mainly from the charged Higgs and chargino
exchange.\\
(ii) The first two generations of up and down-type squarks are almost 
degenerate.\\
 The above features are inferred from the observations noted in Ref. \cite{17}.
 With these simplicities the remaining free parameters are left to be:
 a common mass $m_{\tilde{uL}}$ for the first two generations of the left and
 right-handed squarks; the top squark masses $m_{\tilde{tL}}$, 
 $m_{\tilde{tR}}$ and the top squark mixing angle $\alpha_{\tilde{t}}$;
the masses $M_{\tilde{w}}$ and $M_{H^\pm}$; the $\mu$ and $cot\beta$.

The minimal supersymmetric extension of the standard model gives the new 
contributions to the effective Hamiltonian of eq.(5) by adding some new terms
into the Wilson coefficients. We express the new terms as $C^{SUSY}(\mu)$.
Thus
\begin{equation}
C_i(\mu)=C_i^{SM}(\mu)+C_i^{SUSY}(\mu). 
\end{equation}
\noindent
As in the SM, the Wilson coefficients are calculated in the  scale $\mu=M_W$
at first, then use the renormalization group equation to evolve them down to
the scale $\mu=m_b$. The values of $C_i^{SM}(M_W)$ have been given in 
Ref. \cite{5}.
We calculate the supersymmetric contributions to the Wilson coefficients,
$C_3^{SUSY}$, $C_4^{SUSY}$, $C_5^{SUSY}$, $C_6^{SUSY}$, 
 $C_7^{SUSY}$, $C_8^{SUSY}$, $C_9^{SUSY}$, $C_{10}^{SUSY}$, at $\mu=M_W$ scale.
The authors of Ref. \cite{18} have calculated the contributions of gluino diagrams 
to the process $b\to q'q\bar{q}$. We refer to their results for the gluino
diagram contribution to $C_3^{SUSY}$, $C_4^{SUSY}$, $C_5^{SUSY}$,
$C_6^{SUSY}$ in this work. The initial conditions of the evolution are listed 
in the following,
\begin{eqnarray}
C_1^{SUSY}&=&0,\nonumber\\[4mm]
C_2^{SUSY}&=&0,\nonumber\\[4mm] 
C_3^{SUSY}&=&-\displaystyle\sum_{j=d,s,b}
          \displaystyle\frac{2\alpha_s^2}{g_2^2}
          \displaystyle\frac{M_W^2}{\tilde{m}_g^2}
          \left(s_L(x_j,y)-\frac{1}{6}o_L(x_j,y)\right)
           -\displaystyle\frac{\alpha_s}{24\pi}Z^{SUSY},\nonumber\\[4mm]
C_4^{SUSY}&=&-\displaystyle\sum_{j=d,s,b}
          \displaystyle\frac{\alpha_s^2}{g_2^2}
          \displaystyle\frac{M_W^2}{\tilde{m}_g^2}o_L(x_j,y)+
          \displaystyle\frac{\alpha_s}{8\pi}Z^{SUSY},\\[4mm]
C_5^{SUSY}&=&-\displaystyle\sum_{j=d,s,b}
            \displaystyle\frac{2\alpha_s^2}{g_2^2}
            \displaystyle\frac{M_W^2}{\tilde{m}_g^2}  
             \left(s_R(x_j,y)-\frac{1}{6}o_R(x_j,y)\right)
	     -\displaystyle\frac{\alpha_s}{24\pi}Z^{SUSY},\nonumber\\[4mm]
C_6^{SUSY}&=&-\displaystyle\sum_{j=d,s,b}
            \displaystyle\frac{\alpha_s^2}{g_2^2}
            \displaystyle\frac{M_W^2}{\tilde{m}_g^2}o_R(x_j,y)+
            \displaystyle\frac{\alpha_s}{8\pi}Z^{SUSY},\nonumber
 \end{eqnarray} 
\noindent where \cite{18} 
\begin{eqnarray}
&s_L(x_j,y)=s_R(x_j,y)= \displaystyle\frac{2}{9}g(x_j,y)+
           \displaystyle\frac{4}{9}f(x_j,y),\nonumber\\[4mm]
&o_L=\displaystyle\frac{7}{6}g(x_j,y)-\displaystyle\frac{2}{3}f(x_j,y)+
       \displaystyle\frac{1}{2}F(x_j)+\displaystyle\frac{4}{9}f(x_j),
 \\[4mm]
&o_R=-\displaystyle\frac{1}{3}g(x_j,y)+\displaystyle\frac{7}{3}f(x_j,y)+
    \displaystyle\frac{1}{2}F(x_j)+\displaystyle\frac{4}{9}f(x_j),\nonumber
 \end{eqnarray} 
and
\begin{eqnarray}
f(x,y)&=&\displaystyle\frac{1}{y-x}\left\{\displaystyle\frac{x}{(x-1)^2}lnx-
          \displaystyle\frac{y}{(y-1)^2}lny-
          \displaystyle\frac{1}{x-1}+\displaystyle\frac{1}{y-1}
           \right\},\nonumber\\[4mm]
g(x,y)&=&\displaystyle\frac{1}{x-y}\left\{\displaystyle\frac{x^2}{(x-1)^2}lnx-
         \displaystyle\frac{y^2}{(y-1)^2}lny-\displaystyle\frac{1}{x-1}+
         \displaystyle\frac{1}{y-1}
           \right\},\nonumber\\[4mm]
F(x)&=&\displaystyle\frac{1}{(1-x)^4}\left\{-\displaystyle\frac{3}{2}x^3+
     \displaystyle\frac{15}{2}x^2-\displaystyle\frac{21}{2}x+
     \displaystyle\frac{9}{2}+(2x^3-6x^2+3x+1)lnx\right\},\nonumber\\[4mm]
f(x)&=&\displaystyle\frac{1}{(1-x)^4}\left\{\displaystyle\frac{1}{3}x^3-
   \displaystyle\frac{3}{2}x^2+3x-\displaystyle\frac{11}{6}-lnx\right\},\\[4mm]
x_j&=&\tilde{m}_j^2/\tilde{m}_g^2,~~~~y=\tilde{m}_c^2/\tilde{m}_g^2.\nonumber
\end{eqnarray} 
The values of $C_i^{SUSY}$, ($i=7,\cdots,10$) are
\begin{eqnarray}
C_7^{SUSY}&=&\displaystyle\frac{\alpha}{6\pi}Y_1^{SUSY},\nonumber\\[4mm]
C_8^{SUSY}&=&0,\nonumber\\[4mm]
C_9^{SUSY}&=&\displaystyle\frac{\alpha}{6\pi}Y_2^{SUSY},\\[4mm]
C_{10}^{SUSY}&=&0.\nonumber
\end{eqnarray} 
\noindent And $Z^{SUSY}$, $Y_1^{SUSY}$, $Y_2^{SUSY}$ in eq.(21) and (24) 
are presented in Appendix B.

QCD corrections up to the next-to-leading order are taken into account by
evolving the initial values of the Wilson coefficients down to the 
scale $\mu=m_b$. Because the anomalous dimension matrices of 
the ten operators, $C_i~(i=1,
\cdots,10)$, are the same as in the SM, we can still use the evolution matrix
$\hat{U}(m_b,M_W,\alpha)$ listed in Appendix A. We perform the evolution using
eq.(18) as in section 2). 

After obtaining the Wilson coefficients $C_i~(i=1,\cdots,10)$ with the QCD
corrections up to the next-to-leading order in MSSM, we can still use
eq.(3), (10), (11) to calculate the CP violating parameter with the Wilson
coefficients of the MSSM. We perform those
calculations within the parameter space allowed by experimental limit. For
example, the experiments give the lower bounds of squark and chargino mass are
176 GeV and 45 GeV, respectively\cite{19}. 

Because the complex phase of
$B^0-\bar{B}^0$ mixing is the same as the standard model up to
the minor correction of the order of $(m_c/m_t)^2$ or less\cite{20}, we do not
pay much attention to it here.

We performed our calculation within the very large ranges of all the
parameters ($m_{\tilde{w}}$, $\mu$, $m_{\tilde{g}}$,
$m_{\tilde{u}L}$, $m_{\tilde{t}L}$, $ m_{\tilde{t}R}$, $m_{\tilde{b}}$,
$\alpha_{\tilde{t}}$) which are allowed by the experiment. We find that 
if we keep the Yukawa couplying is perturbative $(cot\beta <3)$\cite{BHP},
the discrepancy from the SM is extremely small, it only happens in the third
desimal number after the zero point.

\begin{center}
{\bf 4) The branching ratio of $\bar{B}^0\to J/\psi K_s$ in the SM, THDM 
and MSSM}
\end{center}

Because there are great theoretical uncertainties in the calculation of the
branching ratio of $\bar{B}^0\to J/\psi K$, in this section we do not want to
give an precise prediction of the branching ratio. We only want to give a
comparison of the predictions of these three models, and to probe the
possibility
to distinguish them by measuring the branching ratio of this decay mode. We
calculate the branching ratio according to the BSW method $\cite{21}$. In our
calculation we take $f_\psi=0.386$GeV, $F_1^{BK}(0)=0.379$ and 
$F_1^{BK}(q^2)=\frac{F_1^{BK}(0)}{1-q^2/m_{1^-}^2}$ where $m_{1^-}=5.43$GeV.
The result of our calculation with the color number being taken as
$N_c=2,~3$ and $\infty$ is that the difference between the predictions of the
SM and MSSM is small, it is only up to a few percent level.

\begin{center}
\section*{III The Study of $\bar{B}^0\to \phi K_s$}
\end{center}

In this section we report the results of our study of $\bar{B}^0\to \phi K_s$
using the effective Hamiltonian given in eq.(5), (14), (18) and (20). Because
there is no tree level diagram contribution to $\bar{B}^0\to \phi K_s$ in the
SM, new physics may have an observable effect. Using the effective
Hamiltonian given in the last section, the amplitude of
$\bar{B}^0\to \phi K_s$ is
\begin{eqnarray}
&\langle K_s\phi|H_{eff}|\bar{B}^0\rangle =\nonumber\\[4mm]
&\displaystyle\frac{G_f}{\sqrt{2}}
       q^*_k\displaystyle\sum_{q=u,c}v_q\left\{
       (1+\displaystyle\frac{1}{N_c})C_3'+
       (1+\displaystyle\frac{1}{N_c})C_4'+C_5'+
       \displaystyle\frac{1}{N_c}C_6'+\displaystyle\frac{3}{2}e_s\left[
       C_7'+\displaystyle\frac{1}{N_c}C_8'\right.\right.\nonumber\\[4mm]
     &  \left.\left.+
       (1+\displaystyle\frac{1}{N_c})C_9'+(\displaystyle\frac{1}{N_c}+1)C_{10}'
       \right]\right\}\langle\phi|(\bar{s}s)_{V-A}|0\rangle
       \langle\bar{K}^0|(\bar{s}b)_{V-A}|\bar{B}^0\rangle
 \end{eqnarray} 
\noindent where $C_i'$ $(i=1,\cdots,10)$ should 
be the renormalization scheme independent
Wilson Coefficients calculated in the SM, THDM and MSSM, respectively.
$e_s=-\frac{1}{3}$ is the charge of the s-quark.

The Decay width of a $\bar{B}^0$ meson at rest decaying into $\phi$ and $K_s$
is 
\begin{center}
\begin{equation}
\Gamma(\bar{B}^0\to\phi K_s)=\frac{1}{8\pi}|\langle\phi K_s|Heff|\bar{B}^0
   \rangle |^2\frac{p}{M_B^2}, 
   \end{equation}
   \end{center}
where
\begin{equation}
|p|=\frac{\{[M_B^2-(M_\phi+M_{K_s})^2][M_B^2-(M_\phi-M_{K_s})^2]\}^{1/2} }
           {2M_B}  
\end{equation}
\noindent is the momentum of $\phi$ or $K_s$. 
The corresponding branching ratio is
given by
\begin{equation}
B_{br}(\bar{B}^0\to\phi K_s)=\frac{\Gamma(\bar{B}^0\to\phi K_s)}
                               {\Gamma_{tot}^{\bar{B}^0} }. 
\end{equation}
We take $\Gamma_{tot}^{\bar{B}^0}=4.22\times 10^{-13}$GeV \cite{19} in our
calculation. The hadronic
matrix element $\langle\phi|(\bar{s}s)_{V-A}|0\rangle \langle\bar{K}^0|
(\bar{s}b)_{V-A}|\bar{B}^0\rangle $ is calculated in the Bauer, Stech, and
Wirbel (BSW) method \cite{21}. The one-body vector matrix element of (V-A)
current is
\begin{center}
\begin{equation}
\langle\phi|(\bar{s}s)_{V-A}|0\rangle=f_{\phi}M_{\phi}\epsilon_{\phi},
\end{equation}
\end{center}
\noindent
where $f_{\phi}=0.233$ GeV, $M_{\phi}=1.02$GeV \cite{19}, $\epsilon_{\phi}$ is
the polarization of $\phi$ meson. The two-body pseudoscalar-pseudoscalar  
matrix element of the vector current is \cite{21}
\begin{equation}
\langle\bar{K}^0|(\bar{s}\gamma_\mu(1-\gamma_5)b|\bar{B}^0\rangle
  =\left(P_B+P_K-\frac{M_B^2-M_K^2}{q^2}q\right)_\mu F_{BK}(q^2,1^-)+
   \frac{M_B^2-M_K^2}{q^2}q_\mu F_{BK}(q^2,0^+), 
\end{equation}
\noindent where $q=P_B-P_{K_s}$,
$F_{BK}(q^2,1^-)=F_{BK}(0,1^-)/(1-\frac{q^2}{M_{1^-}^2})$.
In our calculation, we take $F_{BK}(0,1^-)=0.379$, $M_{1^-}=5.43$GeV \cite{21}.
Now with eq.(3), (25), and (28), we can calculate the CP asymmetry and 
branching ratio of $\bar{B}^0\to \phi K_s$.  Our calculation shows that if we 
take $cot\beta <3$ \cite{BHP}, the difference between the  preictions of
new physics (THDM, MSSM) and SM is too small to be detectable in the future
B factory.

\section*{V. Conclusion and discussion}

We studied the CP asymmetry and the branching ratio of 
$\bar{B}^0\to J/\psi K_s$ and $\bar{B}^0\to \phi K_s$ up to the
leading and next-to-leading order QCD corrections in the standard model, 
the two-Higgs-doublet model and the minimal supersymmetric extension of the
standard model. In our calculation we can neglect the charged Higgs 
tree diagram contribution in the $b\to c\bar{c}s$
decay. Because this diagram gives the contribution to the operator 
$ \bar{c}\gamma_\mu (1+\gamma_5) c\bar{s}\gamma_\mu (1-\gamma_5)b$,
whose coefficient is $-2cot^2\beta \frac{m_c^2}{M_{H^\pm}^2}$, and eq.(19)
means $cot^2\beta \frac{m_c^2}{M_{H^\pm}^2}<(\frac{0.8m_c}{100GeV})^2 $.
This implies the contribution of the charged Higgs 
tree diagram is too small,
and can be safely neglected. So in the two-Higgs-doublet model and the MSSM,
only some loop diagrams are added. Because of the coupling constant 
$\alpha$ and $\alpha_s$ suppression, 
the W-boson tree diagram dominates.
Thus there will be no large difference between the CP asymmetries, branching
ratios of $\bar{B}^0\to J/\psi K_s$ predicted by these three models.

For $\bar{B}^0\to \phi K_s$ decay, new physics only contributes an overall
strong phase, and it does not affect the weak phase much. So in the THDM
amd MSSM, new physics has no large effect in the CP asymmetry of 
$\bar{B}^0\to \phi K_s$. 

After finishing our work, we noticed that the next-to-leading QCD corrections
to $B\to X_s \gamma$ decay are calculated in the THDM in Ref. \cite{Ciu,Bor}.

Finally we come to our conclusion: (i) The tree diagram calculation for
${\cal A}_{cp}$ in $\bar{B}^0\to J/\psi K_s$ is reliable, which is independent 
of the nonfactorization corrections. (ii) It is difficult to
distinguish the SM from the two Higgs-doublet model and MSSM by measuring
CP asymmetries ${\cal A}_{cp}$ and the branching fraction of
 $\bar{B}^0\to J/\psi K_s$ and
$\bar{B}^0\to \phi K_s$ at B factories. 

\vspace{2cm}

\section*{Acknowledgements}

This work was supported in part by the China National Nature Science
Foundation and the Grant of State Commission of Science and Technology 
of China.

\newpage

\begin{center}
{\bf Appendix A}
\end{center}  

The evolution matrix from $M_W$ down to $\mu=m_b$ scale, where we take
$m_b=5.0GeV$.
{\footnotesize
$$\begin{array}{rl}
&\hat{U}(m_b,M_W,\alpha)= \\ 
&\left(
\begin{array}{llllllllll}
   1.115,& -0.246,& 0,&0,&0,&0,&0,&0,&0,&0\\ 
   -0.246,& 1.12,&0,& 0,&0,&0,&0,&0,&0,&0\\
  -0.007,&0.012,&1.10,&-0.203,&0.030,&0.078,& 0.003,&0.008,&-0.012,&0.012\\ 
 0.006,&-0.033,&-0.285,&0.977,&-0.017,&-0.169,&-0.002,&-0.017,& 0.034,&-0.018\\ 
0.003,& 0.009,& 0.035,&0.043,& 0.903,& 0.096,& -0.004,& 0.004,& -0.007,& 0.001\\ 
0.005,&-0.038,&-0.059,& -0.172,& 0.317,&1.717,&-0.003,&-0.029,&0.039,&-0.024\\ 
-0.003,&-0.001,&-0.001,&0.003,&-0.006,&-0.001,&0.913,&0.053,&-0.0096,&-0.004\\ 
-0.0008,&0.0001,&-0.0003,&0.001,&-0.0026,&-0.0104,&0.326,&1.949,&-0.0028,
&-0.0005\\ 
-0.004,&-0.0007,& 0.0038,&0.0016, &-0.0019,& -0.001,& -0.010,
& -0.004,&1.101,&-0.2497\\ 
0.0003,& -0.0001,&-0.001, &0.005,& 0.0002,&-0.0001,& 0.001,& 0.0009,
&-0.244,&1.112
\end{array} \right)
\end{array} 
\eqno(A1)
$$
}
                            
\newpage
\begin{center}
{\bf Appendix B}
\end{center}  

The functions $Z^{SUSY}$, $Y_1^{SUSY}$ and $Y_2^{SUSY}$ appearing in the Wilson  coefficient $C_{7\cdots 9}^{SUSY}$ 
in eq.(21) and (24). Our result is consistent with Ref. \cite{17} 
after replacing the lepton charge with the relevant quark 
charge.

\noindent 1) contribution from charged Higgs loops with a z-boson coupling to
the quark pair:
\begin{eqnarray*}
Y_1&=&\displaystyle\frac{1}{2}cot^2\beta x_t f_1\left(\displaystyle\frac{m_t^2}{m_{H^{\pm}}^2}\right)
         ,\\[4mm]
Y_2&=&-\displaystyle\frac{1}{4}
    \left(\displaystyle\frac{1}{sin^2\theta_w}-2\right)cot^2\beta x_t
       f_1\left(\displaystyle\frac{m_t^2}{m_{H^{\pm}}^2}\right);
   \end{eqnarray*} $$\eqno(B1)$$ 
\noindent 2) contribution from charged Higgs loops with a photon coupling to 
the quark pair:
$$Y_1=Y_2=\displaystyle\frac{1}{18}cot^2\beta 
        f_2\left(\displaystyle\frac{m_t^2}{m_{H^{\pm}}^2}\right).
  \eqno(B2)$$

\noindent 3) contribution from charged Higgs loops with a gluon coupling to 
the quark pair:
$$Z=\displaystyle\frac{1}{6}cot^2\beta 
    f_3\left(\displaystyle\frac{m_t^2}{m_{H^{\pm}}^2}\right).
  \eqno(B3)$$  

\noindent 4) contribution from chargino loops with a z-boson coupling to 
the quark pair:

\begin{eqnarray*}
 Y_1&=&-\displaystyle\frac{2}{g_2^2v_{ts}^*v_{tb}}\displaystyle\sum_{A,B=1}^6
       \displaystyle\sum_{I,J=1}^2(X_I^{U_L})_{2A}^+(X_J^{U_L})_{B3}
    \left\{c_2(m^2_{\tilde{\chi}_I^{\pm}},m^2_{\tilde{u}_A},m^2_{\tilde{u}_B})
    (\Gamma^{U_L}\Gamma^{U_L+})_{AB}\delta_{IJ}\right.\\[4mm]
 &&-c_2(m^2_{\tilde{u}_A},m^2_{\tilde{\chi}_I^{\pm}},m^2_{\tilde{\chi}_J^{\pm}})
   \delta_{AB}V_{I1}^*V_{J1}+\displaystyle\frac{1}{2}m_{\tilde{\chi}_I^{\pm}}
   m_{\tilde{\chi}_J^{\pm}}
   c_0(m^2_{\tilde{u}_A},m^2_{\tilde{\chi}_I^{\pm}},m^2_{\tilde{\chi}_J^{\pm}})
   \delta_{AB}U_{I1}U_{J1}^*\left.\right\},\\[4mm]
 Y_2&=&\left(\displaystyle\frac{1}{sin^2\theta_w}-2\right)
     \displaystyle\frac{1}{g_2^2v_{ts}^*v_{tb}}\displaystyle\sum_{A,B=1}^6
       \displaystyle\sum_{I,J=1}^2(X_I^{U_L})_{2A}^+(X_J^{U_L})_{B3}
    \left\{c_2(m^2_{\tilde{\chi}_I^{\pm}},m^2_{\tilde{u}_A},m^2_{\tilde{u}_B})
    (\Gamma^{U_L}\Gamma^{U_L+})_{AB}\delta_{IJ}\right.\\[4mm]
 &&-c_2(m^2_{\tilde{u}_A},m^2_{\tilde{\chi}_I^{\pm}},m^2_{\tilde{\chi}_J^{\pm}})
   \delta_{AB}V_{I1}^*V_{J1}+\displaystyle\frac{1}{2}m_{\tilde{\chi}_I^{\pm}}
   m_{\tilde{\chi}_J^{\pm}}
   c_0(m^2_{\tilde{u}_A},m^2_{\tilde{\chi}_I^{\pm}},m^2_{\tilde{\chi}_J^{\pm}})
   \delta_{AB}U_{I1}U_{J1}^*\left.\right\}.
\end{eqnarray*} $$\eqno(B4)$$
\noindent 5) contribution from chargino loops with a photon coupling to 
the quark pair:
$$Y_1=Y_2=-\displaystyle\frac{1}{9g_2^2v_{ts}^*v_{tb}}\displaystyle\sum_{A=1}^6
     \displaystyle\sum_{I=1}^2\displaystyle\frac{m^2_w}{m^2_{\tilde{u}_A}}
     (X_I^{U_L})_{2A}^+(X_I^{U_L})_{A3}
     f_4\left(\displaystyle\frac{m^2_{\tilde{\chi}_I^{\pm}}}
     {m^2_{\tilde{u}_A}}\right). \eqno(B5)$$
\noindent 6) contribution from chargino loops with a gluon coupling to 
the quark pair:
$$Z=-\displaystyle\frac{1}{3g_2^2v_{ts}^*v_{tb}}\displaystyle\sum_{A=1}^6
     \displaystyle\sum_{I=1}^2\displaystyle\frac{m^2_w}{m^2_{\tilde{u}_A}}
     (X_I^{U_L})_{2A}^+(X_I^{U_L})_{A3}
     f_5\left(\displaystyle\frac{m^2_{\tilde{\chi}_I^{\pm}}}
     {m^2_{\tilde{u}_A}}\right). \eqno(B6)$$

\noindent 7) contribution from neutralino loops with a z-boson coupling to
the quark pair:
 \begin{eqnarray*}
 Y_1&=&-\displaystyle\frac{2}{g_2^2v_{ts}^*v_{tb}}\displaystyle\sum_{A,B=1}^6
       \displaystyle\sum_{I,J=1}^4(Z_I^{D_L})_{2A}^+(Z_J^{D_L})_{B3}
    \left\{c_2(m^2_{\tilde{\chi}_I^0},m^2_{\tilde{d}_A},m^2_{\tilde{d}_B})
    (\Gamma^{D_R}\Gamma^{D_R+})_{AB}\delta_{IJ}\right.\\[4mm]
 &&-c_2(m^2_{\tilde{d}_A},m^2_{\tilde{\chi}_I^0},m^2_{\tilde{\chi}_J^0})
   \delta_{AB}(N_{I3}^*N_{J3}-N_{I4}^*N_{J4})-
   \displaystyle\frac{1}{2}m_{\tilde{\chi}_I^0}m_{\tilde{\chi}_J^0}
   c_0(m^2_{\tilde{d}_A},m^2_{\tilde{\chi}_I^0},m^2_{\tilde{\chi}_J^0})
   \delta_{AB}(N_{I3}N_{J3}^*\\[4mm]
   &&-N_{I4}N_{J4}^*)\left.\right\},\\[4mm]
 Y_2&=&\left(\displaystyle\frac{1}{sin^2\theta_w}-2\right)
   \displaystyle\frac{1}{g_2^2v_{ts}^*v_{tb}}\displaystyle\sum_{A,B=1}^6
       \displaystyle\sum_{I,J=1}^4(Z_I^{D_L})_{2A}^+(Z_J^{D_L})_{B3}
    \left\{c_2(m^2_{\tilde{\chi}_I^0},m^2_{\tilde{d}_A},m^2_{\tilde{d}_B})
    (\Gamma^{D_R}\Gamma^{D_R+})_{AB}\delta_{IJ}\right.\\[4mm]
 &&-c_2(m^2_{\tilde{d}_A},m^2_{\tilde{\chi}_I^0},m^2_{\tilde{\chi}_J^0})
   \delta_{AB}(N_{I3}^*N_{J3}-N_{I4}^*N_{J4})-
   \displaystyle\frac{1}{2}m_{\tilde{\chi}_I^0}m_{\tilde{\chi}_J^0}
   c_0(m^2_{\tilde{d}_A},m^2_{\tilde{\chi}_I^0},m^2_{\tilde{\chi}_J^0})
   \delta_{AB}(N_{I3}N_{J3}^*\\[4mm]
   &&-N_{I4}N_{J4}^*)\left.\right\}.
   \end{eqnarray*} $$\eqno(B7)$$ 
\noindent 8) contribution from neutralino loops with a photon coupling to 
the quark pair:
$$Y_1=Y_2=\displaystyle\frac{1}{54g_2^2v_{ts}^*v_{tb}}\displaystyle\sum_{A=1}^6
     \displaystyle\sum_{I=1}^4\displaystyle\frac{m^2_w}{m^2_{\tilde{d}_A}}
     (Z_I^{D_L})_{2A}^+(Z_I^{U_L})_{A3}
     f_6\left(\displaystyle\frac{m^2_{\tilde{\chi}_I^0}}
     {m^2_{\tilde{d}_A}}\right). \eqno(B8)$$
\noindent 9) contribution from gluino loops with a z-boson coupling to
the quark pair:
\begin{eqnarray*}
 Y_1&=&-\displaystyle\frac{16g_3^2}{3g_2^2v_{ts}^*v_{tb}}
 \displaystyle\sum_{A,B=1}^6
 (\Gamma^{D_L})^+_{2A}(\Gamma^{D_L})_{B3}
c_2(m^2_{\tilde{g}},m^2_{\tilde{d}_A},m^2_{\tilde{d}_B})
(\Gamma^{D_R}\Gamma^{D_R+})_{AB},\\[4mm]
 Y_2&=&\displaystyle\frac{8g_3^2}{3g_2^2v_{ts}^*v_{tb}}
 \left(\displaystyle\frac{1}{sin^2\theta_w}-2\right)
 \displaystyle\sum_{A,B=1}^6
 (\Gamma^{D_L})^+_{2A}(\Gamma^{D_L})_{B3}
c_2(m^2_{\tilde{g}},m^2_{\tilde{d}_A},m^2_{\tilde{d}_B})
(\Gamma^{D_R}\Gamma^{D_R+})_{AB}.
\end{eqnarray*} $$\eqno(B9)$$
\noindent 10) contribution from gluino loops with a photon coupling to 
the quark pair:
$$Y_1=Y_2=\displaystyle\frac{4g_3^2}{81g_2^2v_{ts}^*v_{tb}}
     \displaystyle\sum_{A=1}^6
     \displaystyle\frac{m^2_w}{m^2_{\tilde{d}_A}}
 (\Gamma_I^{D_L})_{2A}^+(\Gamma_I^{U_L})_{A3}
 f_6\left(\displaystyle\frac{m^2_{\tilde{g}}}
     {m^2_{\tilde{d}_A}}\right). \eqno(B10)$$
     
\noindent The functions appearing in the eq.(B1)$\cdots$(B10) are given by
\begin{eqnarray*}
f_1(x)&=&\displaystyle\frac{x}{1-x}+
    \displaystyle\frac{x}{(1-x)^2}lnx,\\[4mm]
f_2(x)&=&\displaystyle\frac{38x-79x^2+47x^3}{6(1-x)^3}+
    \displaystyle\frac{4x-6x^2+3x^4}{(1-x)^4}lnx,\\[4mm]
f_3(x)&=&\displaystyle\frac{16x-29x^2+7x^3}{6(1-x)^3}+
        \displaystyle\frac{2x-3x^2}{(1-x)^4}lnx,\\[4mm]             
f_4(x)&=&\displaystyle\frac{52-101x+43x^2}{6(1-x)^3}+
      \displaystyle\frac{6-9x+2x^3}{(1-x)^4}lnx,\\[4mm]     
f_5(x)&=&\displaystyle\frac{2-7x+11x^2}{6(1-x)^3}+
         \displaystyle\frac{x^3}{(1-x)^4}lnx,\\[4mm]
f_6(x)&=&\displaystyle\frac{2-7x+11x^2}{(1-x)^3}+
  \displaystyle\frac{6x^3}{(1-x)^4}lnx,\\[4mm]
c_0(m_1^2,m_2^2,m_3^2)&=&-\left(
     \displaystyle\frac{m_1^2ln(m_1^2/\mu^2)}{(m_1^2-m_2^2)
      (m_1^2-m_3^2)}+(m_1\leftrightarrow m_2)+(m_1\leftrightarrow m_3)\right),
       \\[4mm]
c_2(m_1^2,m_2^2,m_3^2)&=&\displaystyle\frac{3}{8}-
      \displaystyle\frac{1}{4}\left(
      \displaystyle\frac{m_1^4ln(m_1^2/\mu^2)}{(m_1^2-m_2^2)
      (m_1^2-m_3^2)}+(m_1\leftrightarrow m_2)+(m_1\leftrightarrow m_3)\right).
\end{eqnarray*} 
\noindent The symbol conventions in eq.(B1$\cdots$ B10) are the same as in
 Ref. \cite{17}.       

\newpage

\begin{center}
{\bf Figure Captions}
\end{center}

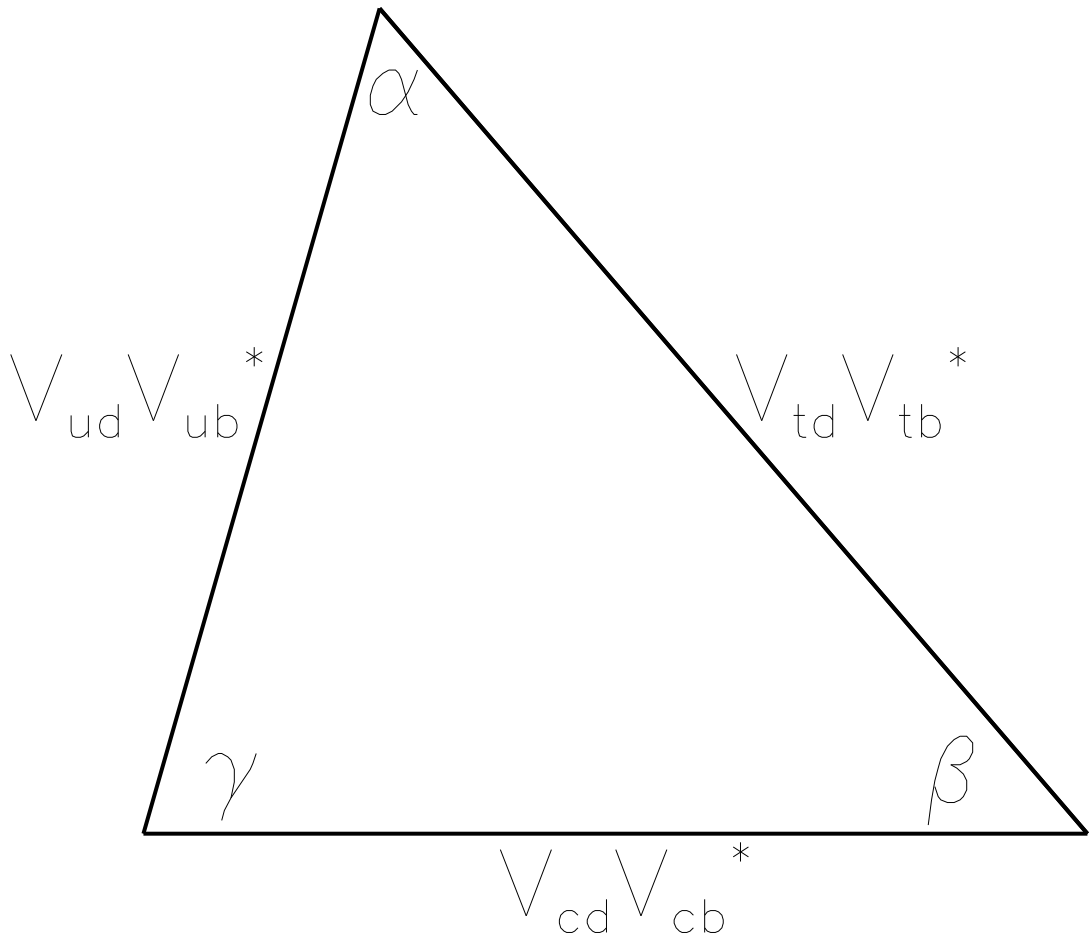
Fig.1: The unitarity triangle of CKM matrix in the complex plane.

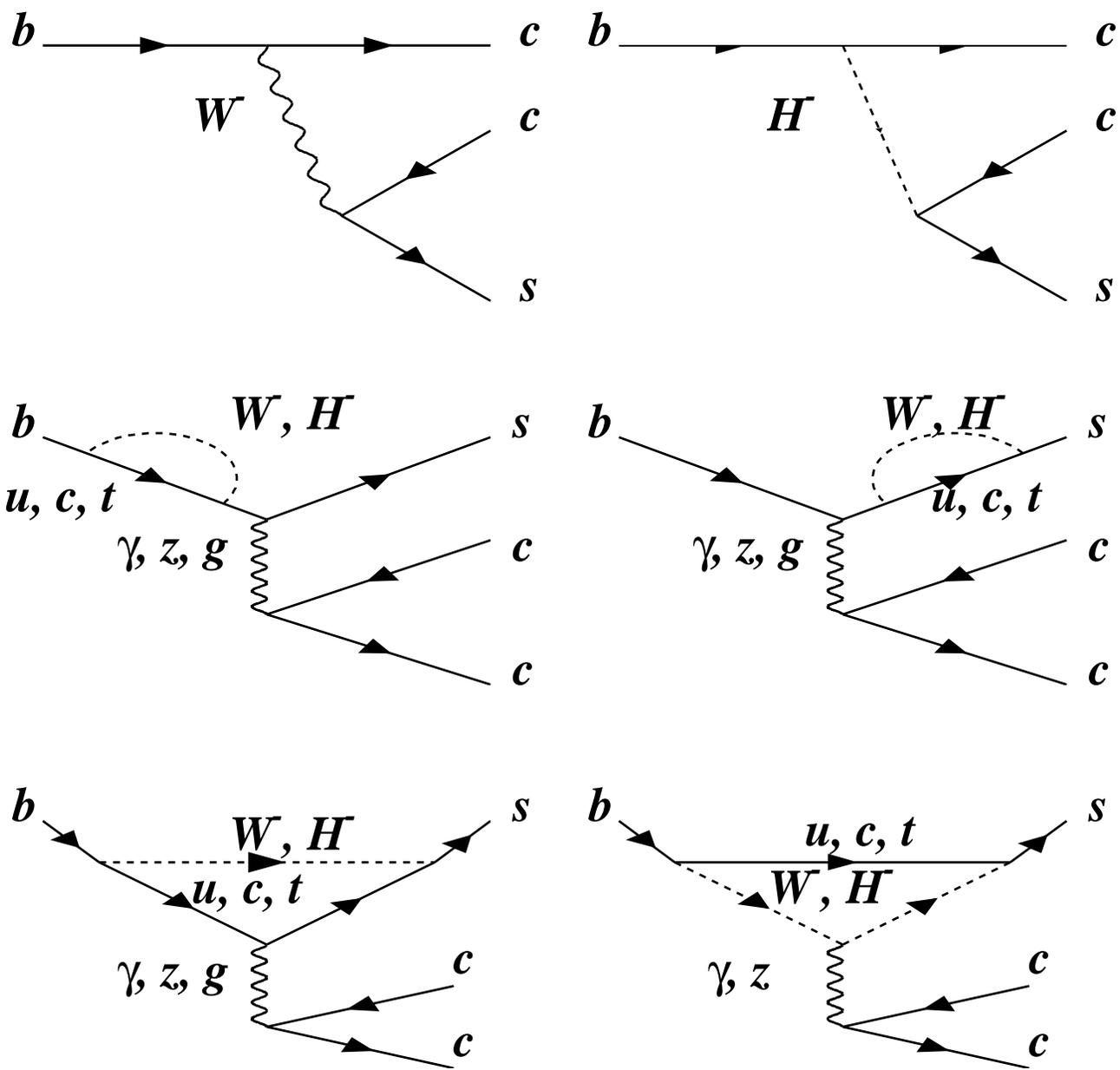
Fig.2: Diagrams contributing to the process $b\to c\bar{c} s$ in the 
       two-Higgs-doublet model, which mediate $\bar{B}^0\to J/\psi K_s$
       decay.

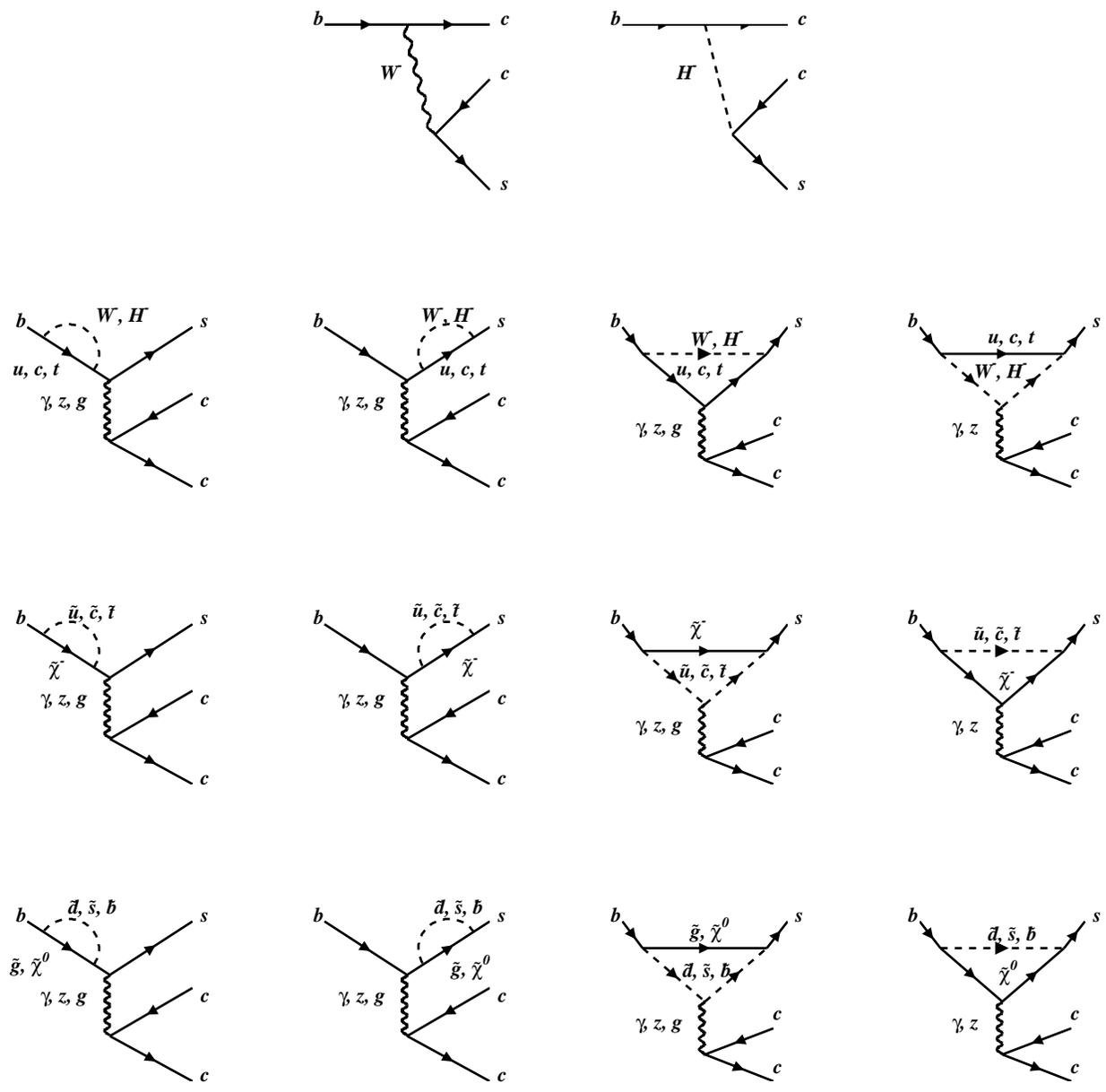
Fig.3: Diagrams contributing to the process $b\to c\bar{c} s$ in the MSSM.

\end{document}